\documentclass[10pt]{article}
\usepackage{graphicx}
\usepackage{bm}
 \usepackage{amsfonts,amssymb}
\usepackage{amsmath}

\numberwithin{equation}{section}

\def\a{\alpha}
\def\b{\beta}
\def\g{\gamma}
\def\d{\delta}
\def\ep{\varepsilon}

\def\vp{\varphi}

\def\a{\alpha}
\def\b{\beta}

\def\g{\gamma}
\def\d{\delta}
\def\g{\gamma}

\def\J{{\mathcal{J}}}

\def\F{{\cal{F}}}
 \def\H{{\mathcal{H}}}
  \def\D{{\mathcal{D}}}

 \def\K{{\mathcal{K}}}
 
\begin{document}

 \title{Covariant jump conditions in electromagnetism}
\author{Yakov Itin\\
Institute of Mathematics, Hebrew University of
   Jerusalem \\
   and Jerusalem College of Technology\\
  email: itin@math.huji.ac.il}
\maketitle






%

\begin{abstract}
A generally covariant four-dimensional representation of Maxwell's electrodynamics in a generic material medium can be achieved straightforwardly in the metric-free formulation of electromagnetism.
In this setup, the electromagnetic phenomena  described by two tensor fields, which satisfy Maxwell's equations.
  A generic tensorial constitutive relation between these fields  is an independent ingredient of the theory. By use of different constitutive relations (local and non-local, linear and non-linear, etc.), a wide area of applications can be covered. 
In the current paper, we present  the jump conditions for the fields and for the energy-momentum tensor on an arbitrarily moving surface between two  media.  From the differential and integral Maxwell equations, we derive  the covariant   boundary conditions, which are independent of  any metric and connection. These conditions include the covariantly defined surface current and are applicable to an arbitrarily moving smooth curved boundary surface. As an application of the presented jump formulas, we derive a Lorentzian type metric as a condition for existence the wave front in isotropic media. This result holds for the ordinary materials as well as for the metamaterials with the negative material constants.  
\end{abstract}

\section{Introduction}
Maxwell's system of the electromagnetic field equations was initially formulated  as a collection of the   experimentally established electromagnetic laws, i.e., as a phenomeno\-logical model.     Eventually, this model was recognized   as a Lorentz covariant and even general relativistic covariant theory, see \cite{Zom},\cite{Pauli},\cite{LL},\cite{Jack}.  The rich group of transformations  associated with  the Maxwell system is  not only an aesthetic feature. In fact, the general covariant form of the  equations is very useful even on a flat spacetime when  curvilinear coordinates are used. It becomes  compulsory  when  the  contribution of the gravitational field to electromagnetism is considered. Moreover, the generally covariant form can serve as a guiding framework  for possible theoretical extensions of the classical electromagnetic theory. 

The procedure of the reformulation of the Maxwell system into a covariant form requires a proper tensorial redefinition of the basic electromagnetic variables. It is quite well known from Lorentz, Minkowski, and Einstein. A historical account of this development  was given  recently in \cite{ann-Hehl} and \cite{ann-Damour}.  

  Another principal feature of Maxwell's system is that it is not dependent  from the geometry of the  underlying  manifold, neither from the metric  nor from the connection.  This fact was recognized long ago \cite{Kottler},\cite{Cartan},\cite{Dantzig},\cite{Post}, but its comprehensive treating was provided only recently, see  \cite{birkbook} and the references given therein. In vacuum,  the metric-independence of the Maxwell system is just a nice mathematical property, but in electromagnetism of media this fact has a firm physical meaning. Indeed, the only way one can observe the spacetime metric in a non-trivial electromagnetic medium is by studying the propagation of light in it.
 However, the light propagation is governed by the electromagnetic moduli of the medium, not by the Lorentzian metric. In general, the corresponding optical geometry is not Lorentzian. Instead, it obeys a Finslerian geometry \cite{Perlick}.  Thus, the Lorentzian metric tensor of the vacuum geometry cannot actually be observed inside of  electromagnetic media. Under these circumstances, the metric tensor turns out to be a redundant notion that must be omitted from our description. In fact,  the metric-free formulation  is acceptable in a vast range of the  pure electromagnetic problems, especially in optics. Actually, when a proper definition of the variables is used, the metric tensor ``cancels out``  from all equations. Some examples of  such a metric-free representation were derived recently:
(i)  The generalized covariant dispersion relation \cite{birkbook}, \cite{Itin:2009aa}. 
(ii)  The generalized covariant photon propagator \cite{Itin:2007av}.
(iii) The non-relativistic dispersion relation for anisotropic media \cite{Itin:pla}. 
Particularly, the axion, skewon and dilaton partners of the photon naturally emerge in this framework. The generic metric-free construction also can serve as a working model to organize the experimental electromagnetic data in order to estimate the possible deviations from the standard Maxwell theory \cite{Lam}, \cite{Kostel}.  

In the current paper, we turn to an additional  ingredient of the Maxwell theory --- {\it the jump conditions} for the fields and for the energy-momentum tensor on an arbitrarily moving surface between two different media.  These relations are usually presented in a non-relativistic three-dimensional form with an explicit use of the Euclidean metric. Our aim is to derive the metric-free and general relativistic covariant form of the jump  conditions. This question was studied rather intensively, see \cite{nasa}, \cite{Rawson-Harris}, \cite{Yeh},  \cite{Gur}, \cite{Lindell}, \cite{PH}. However, to our knowledge, a manifestly  covariant and metric-free result was not yet presented in the literature. 

The organization of the paper is as follows: In the next section, we recall the  covariant metric-free formulation of the Maxwell system. The third section is devoted to a derivation of the covariant jump conditions for electromagnetic field on an arbitrary moving  surface. We start with the known static jump  conditions and show how they can be rewritten in the metric-free form. A complete description of the field jumps must include surface electric charges and currents. The covariant metric-free description of the surface quantities dictates the use of  the Dirac $\d$-form instead of the Dirac $\d$-function. The covariant meaning of the boundary hypersurface in the four-dimensional manifold is discussed  and the covariant metric-free definition of the  surface electric current  introduced. First, we derive the metric-free covariant jump conditions from the Maxwell equations by formal differentiation of the  Heaviside step-distribution. In parallel with that, we present a derivation of the same jump conditions from the integral Maxwell equations. 
We study the $(1+3)$-decomposition of the covariant jump  conditions.   In the static case,   the classical jump conditions are reinstated. 
We also discuss the  number of the independent boundary conditions. Their number turns out to be  six, both in the covariant description and in the $(1+3)$ decomposition. 
An example in section 8 represents the use of the covariant metric-free jump conditions in the case of a uniform isotropic media. The consistence of the jump conditions induce the possibility for the existence of shock waves in this medium.  Their front has a Lorentzian-type gradient with a speed of $v=1/\sqrt{\varepsilon\mu}$ instead  of the  vacuum speed of light. Thus, the Lorentzian-type metric is obtained in the initially metric-free formulation.

We complete the paper with a discussion of the results.  

\section{Maxwell equations}
In this section, we give a brief account of the covariant metric-free  presentation of the  electromagnetism in media. For a comprehensive description, see  \cite{birkbook} and the references given therein. 
Let us start with a differential forms formulation of the Maxwell equations. On an arbitrary 4-dimensional differential manifold, the two field equations read
\begin{equation}\label{curr-1}
dF=0\,,\qquad d\H=\J\,.
\end{equation}
Here $F$ is an untwisted 2-form of  the {\it field strength}, $\H$ is a twisted 2-form of the {\it field excitation},    and $\J$ is a twisted 3-form of the {\it electric current}. 
In an integral formulation, the equations (\ref{curr-1}) are represented by a system
\begin{equation}\label{int1}
\int_{\partial \Sigma} F=0\,,\qquad   \int_{\partial \Sigma}\H=\int_{\Sigma}\J\,,
\end{equation}
where $\Sigma$ is an arbitrary connected 3-dimensional domain bounded by the closed 2-dimensional surface $\partial \Sigma$. In fact, one can interpret these equations as the expressions of the magnetic flux conservation law and the electric charge conservation law,  respectively. 

The systems (\ref{curr-1}) and (\ref{int1}) are explicitly general relativistic covariant and do not require a metric for their formulation.  
In a coordinate basis, the differential forms can be expanded as
\begin{equation}\label{forms}
F=\frac 12F_{ij}dx^i\wedge dx^j\,,\qquad \H=\frac 12 \H^{ij}\ep_{ijkl}dx^k\wedge dx^l\,.
\end{equation}
Here, the spacetime coordinates $\{x^0,x^1,x^2,x^3\}=\{t,x,y,z\}$ are used. The Roman indices   run over the range of $i,j,\cdots=0,1,2,3$. 
We use the 4-dimensional Levi-Civita permutation pseudo-tensor $\epsilon^{ijkl}$ which is normalized   
as $\epsilon^{0123}=1$. 
The odd 3-form of the electric current is expanded as  
 \begin{equation}\label{forms-1}
\J=\frac 16 \J^{i}\ep_{ijkl}dx^j\wedge dx^k\wedge dx^l\,.
\end{equation}

With the definitions above  the tensorial form  of Maxwell's equations  is given by 
\begin{equation}\label{gen-1}
\ep^{ijkl}F_{jk,l}=0\,, \qquad \H^{ij}{}_{,i}= \J^j\,.
\end{equation}

The $(1+3)$-decomposition of the field tensors  $F_{ij}$ and  $\H^{ij}$ reads 
 \begin{eqnarray}\label{vec-ten3}
  F_{0\a}=E_\a\,,\qquad&& F_{\a\b}=-\ep_{\a\b\g}B^\g\,,\\
  \label{vec-ten4}
 \H^{0\a}= -D^\a\,,\qquad&&\H^{\a\b}=\ep^{\a\b\g}H_\g\,,
  \end{eqnarray}
where the Greek indices run over the range $\a,\b=1,2,3$. 
 The inverse relations are given by    
\begin{eqnarray}\label{vec-ten1}
  E_\a=F_{0\a}\,,\qquad&& B^\a=-\frac 12 \ep^{\a\b\g}F_{\b\g}\,,\\
  \label{vec-ten2}
  D^\a=-\H^{0\a}\,,\qquad&&H_\a=\frac 12 \ep_{\a\b\g}\H^{\b\g}\,.
  \end{eqnarray}
The correspondence between these two sets of formulas can be checked by use of the relations $\ep_{\a\b\g}\ep^{\a\mu\nu}=\d^\mu_\b\d^\nu_\g-\d^\nu_\b\d^\mu_\g$ and $\ep_{\a\b\g}\ep^{\a\b\mu}=2\d^\mu_\g$. 
The $(1+3)$-decomposition of the electric current $J^i$ is given by 
\begin{equation}\label{curr}
\J^0=\rho\,,\qquad \J^\a= j^\a\,.
\end{equation}

When these definitions are substituted into (\ref{gen-1}), we come to the  ordinary 3D-representation of Maxwell's equations in media, see \cite{Jack}, 
  \begin{eqnarray}\label{max-1}
  \nabla\cdot\,{\mathbf B}=0\,,\qquad&& \nabla\times\,{\mathbf E} +\frac{\partial{\mathbf B}}{\partial t}=0\,,\\
  \label{max-1a}
 \nabla\cdot\,{\mathbf D}=\rho\,,\qquad&& \nabla\times\,{\mathbf H} - \frac{\partial{\mathbf D}}{\partial t}= {\mathbf j}\,.
  \end{eqnarray}
The definitions for the vector operations are assumed here  as
\begin{equation}\label{vec-1}
\nabla\cdot\,{\mathbf B}=B^\a{}_{,\a}\,, \qquad \nabla\cdot\,{\mathbf D}=D^\a{}_{,\a}\,,
\end{equation}
and
\begin{equation}\label{vec-2}
\nabla\times\,{\mathbf H}=-\ep^{\a\b\g}H_{\b,\g}\,, \qquad \nabla\times\,{\mathbf E}=-\ep^{\a\b\g}E_{\b,\g}\,,
\end{equation}
where $\epsilon^{\a\b\g}$ denotes the three-dimensional Levi-Civita's permutation pseudo-tensor with $\epsilon^{123}=1$.

It is an extremely   significant  fact that the 3-dimensional system (\ref{max-1}-\ref{max-1a})  is already general relativistic invariant and metric-free.  Indeed, even being written in the usual three-dimensional notations it is equivalent to the invariant system  (\ref{curr-1}) and independent of a metric structure.

\section{Jump conditions}
 
In this section, we  start with the standard textbook boundary conditions, which are the consequences of Maxwell's field equations. These conditions include 3-component vector fields depending of time and position. Certainly, they are not preserved under 4-dimensional transformations. Moreover, the conditions  deal with a static flat boundary  which is also a non-covariant notion. Furthermore,  these expressions  use explicitly the Euclidean metric of the three-dimensional position space. Observe that the underlying geometric structure is ill defined. Due to the independent free variables, the space must be 4-dimensional. The fields, however, are assumed to be described by 3-vectors. 

Our aim is to remove all these restrictions in order  
to obtain the covariant metric-free 4-dimensional boundary conditions.  

\subsection{Static jump conditions}
Let us first consider  a non-moving surface between two media.  In a chosen system of Cartesian spatial coordinates $\{x,y,z\}$, a static surface can  be represented by an implicit equation 
\begin{equation}\label{bound-01}
\vp(x,y,z)=0\qquad {\rm -\qquad surface.}
\end{equation}
We assume  $\vp(x,y,z)$ to be a smooth real function defined in the whole space, i.e., $\vp:{\mathbb R}^3\to\mathbb R$. For an arbitrary point $(x,y,z)$ in the medium, we assume one of the conditions 
\begin{equation}\label{bound-02}
\vp(x,y,z)<0\qquad {\rm -\qquad  medium\,\, (1),}
\end{equation}
 or, 
\begin{equation}\label{bound-03}
\vp(x,y,z)>0\qquad {\rm -\qquad medium \,\,(2).}
\end{equation}
For an arbitrary  physical variable $G=G(t,x,y,z)$, we describe its  discontinuity at the boundary $\vp(x^\a)$ by  the quantity 
\begin{equation}\label{bound-04}
\big[G\big]=\lim_{\vp\to 0^+}G(t,x,y,z)\,-\lim_{\vp\to 0^-}G(t,x,y,z)\,.  
\end{equation}
For a static boundary surface,  one introduces a normal unit vector 
\begin{equation}\label{nabla}
\mathbf n=\frac{\nabla \vp}{||\nabla \vp||}\,
\end{equation}
and writes the standard static 3-dimensional boundary conditions as  \cite{Jack}
\begin{eqnarray}\label{bound-1}
{\mathbf n}\cdot\big[\mathbf B\big]=0\,,\qquad&& {\mathbf n}\times
\big[\mathbf E\big]=0\,,\\
  \label{bound-1a}
{\mathbf n}\cdot\big[\mathbf D\big]=\sigma\,,\qquad&& {\mathbf n}\times\big[\mathbf H\big]={\mathbf K}\,.
  \end{eqnarray}
Here $\sigma$ and ${\mathbf K}$ are the 3-dimensional surface charge scalar and  surface current vector respectively.   At first glance, the conditions (\ref{bound-1}--\ref{bound-1a})  depend  crucially on the Euclidean space metric. Indeed, the 3-dimensional metric is involved in the definition of the unit normal ${\mathbf n}$ and of the scalar (dot)  product. We will see that these equations can be readily reformulated in a metric-free form. 

First, we substitute  (\ref{nabla}) into (\ref{bound-1}--\ref{bound-1a}) to obtain 
\begin{eqnarray}\label{bound-1x}
\nabla \vp\cdot\big[\mathbf B\big]=0\,,\qquad&& \nabla \vp\times
\big[\mathbf E\big]=0\,,\\
  \label{bound-1ax}
\nabla \vp\cdot\big[\mathbf D\big]=\sigma ||\nabla \vp||\,,\qquad&& \nabla \vp\times\big[\mathbf H\big]={\mathbf K} ||\nabla \vp||\,.
  \end{eqnarray}
The scalar product terms appearing in these equations can be written  as a contraction of a covector $\nabla \vp$ with the vector fields  $\mathbf B=B^\a\partial_\a$ and $\mathbf D=D^\a\partial_\a$. Note that such a description of the fields  $\mathbf B$ and $\mathbf D$ as contravariant vectors is necessary for the covariant meaning of the field equations. Alternatively, $\nabla  \vp$ comes as a covariant vector (or covector). 
Consequently, we have the identifications  
\begin{equation}
\nabla \vp\cdot\mathbf B=\vp_{,\a} B^\a\,,\qquad 
\nabla \vp\cdot\mathbf D=\vp_{,\a} D^\a\,.
\end{equation}
 The standard expression for the cross-product also can be assembled in a non-metrical form if the corresponding fields are treated as down-indexed, i.e., as covectors.   
As we have already seen, this is the case with the fields $\mathbf E$ and $\mathbf H$. Thus,  we have a metric-free representation
\begin{equation}
\left(\nabla \vp\times{\mathbf E}\right)^\a=\ep^{\a\b\g}\vp_{,\b}E_\g\,,\qquad \left(\nabla \vp\times{\mathbf H}\right)^\a=\ep^{\a\b\g}\vp_{,\b}H_\g\,.
\end{equation}
Consequently,  the system (\ref{bound-1}-\ref{bound-1a}) takes the form 
\begin{eqnarray}\label{bound-new2}
\vp_{,\a}\big[ B^\a\big]=0\,,\qquad&& \ep^{\a\b\g}\vp_{,\b}
\big[ E_\g\big]=0\,,\\
  \label{bound-new2x}
\vp_{,\a}\big[ D^\a\big]=\sigma ||\nabla \vp||\,,\qquad&& \ep^{\a\b\g}\vp_{,\b}\big[ H_\g\big] ={\K^\a}||\nabla \vp||\,.
\end{eqnarray}
Substituting (\ref{vec-ten1}), we can rewrite this system via the components of the 4-dimensional tensors as 
\begin{eqnarray}\label{bound-new2a}
\ep^{\a\b\g}\vp_\a \big[F_{\b\g}\big]=0\,,\qquad && \ep^{\a\b\g}\vp_\b \big[F_{0\g}\big]=0\,,\\
  \label{bound-new2b}
  \vp_\a\big[\H^{\a0}\big]=\sigma ||\nabla \vp||\,,\qquad&& 
  \vp_\b\big[\H^{\a\b}\big]={\K^\a}||\nabla \vp||\,.
\end{eqnarray}
Now the metric tensor remains only in the definition of the norm $ ||\nabla \vp||$. It will be removed in the sequel   by a suitable redefinition of the surface quantities. 
Note that the system of these equations is still 3-dimensional and the boundary between two media is still static.  

\subsection{Moving boundary }
A covariant definition of a boundary between two media by necessity must be time-dependent. 
On a four-dimensional manifold, the boundary spans a three-dimensional hypersurface. The 
 general representation of such a hypersurface is given   by a scalar equation 
\begin{equation}\label{bound-2-1}
\vp(x^i):=\vp(t,x,y,z)=0\qquad {\rm -\qquad surface.}
\end{equation}
This equation can describe a submanifold of a rather complicated topology, for instance, with  self-intersections. 
In order to remove these cases, it is enough to require the gradient of $\vp$ to be non-zero at all points where (\ref{bound-2-1}) holds. 

Moreover, different functional dependence of $\vp$ on the  time coordinate can be used to describe the same boundary.   It should be noted that, for a given hypersurface, a choice of a function $\vp(x^i)$ is highly non-unique. Indeed, for  an arbitrary monotonic function $f$ with $f(0)=0$, two  equations $f(\vp)=0$ and $\vp=0$
represent the same hypersurface. Moreover, the physical dimensions of the quantities $\vp$ and $f(\vp)$ can be different. We will take care of these difficulties  subsequently.  

 Not all of the surfaces (\ref{bound-2-1}) can serve as a physical boundary between two media.  We restrict ourself to a small piece of the surface which is embedded without self-intersections in that part of the $4D$ manifold where the coordinates $x^i$ are defined. Global embedding  problems will not play any role in our considerations. We assume  that $\vp(t,x,y,z)$ is a smooth function in the whole domain.   Moreover, we restrict our considerations to a small region which is separated by the surface (\ref{bound-2-1}) into two disjoint pieces  representing two media.  
Consequently, for an arbitrary point  into the first medium, we assume 
\begin{equation}\label{bound-2-2}
\vp(t,x,y,z)<0\qquad {\rm -\qquad medium \, (1),}
\end{equation}
 while, for a point into the second medium, we have
\begin{equation}\label{bound-2-3}
\vp(t,x,y,z)>0\qquad {\rm -\qquad medium \, (2).}
\end{equation}
Denote the 4-dimensional gradient of $\vp(x^i)$ as
\begin{equation}\label{bound-2x}
\vp_{,i}=\frac{\partial \varphi}{\partial x^i}\,
\end{equation}
and assume it to be defined and non-zero in the whole   spacetime domain. 
Observe that this quantity  naturally comes with a lower index.  Hence, it  must be treated as a 4-covector.  
Note that $\vp_{,i}$ does not depend of  the metric tensor and it cannot be treated as   a normal    to the surface (the orthogonality  notion is not defined in the metric-free context).    
 For a static boundary, the identification $\vp_{,i}=(0,{\mathbf n})$ can be assumed.   
  In the notation of differential forms, we have an even 1-form 
\begin{equation}\label{bound-2xx}
d\vp=\vp_{,i} dx^i\,,
\end{equation}
whose components are the same as in (\ref{bound-2x}).

\subsection{Surface current}
For a fully covariant boundary condition, we need the notion of a covariant   surface current. 
We treat the  electric surface current as a special type of an ordinary electric current which can be substituted into the right-hand side of the  inhomogeneous  Maxwell  equation and if  integrated over a 3-dimensional surface, yields the  total charge.  Consequently, the four-dimensional surface current must be represented by a differential 3-form with a support on the surface. To deal with such singular currents, which are  localized only  on  a surface $\vp(x^i)=0$, we define an even  1-form 
\begin{equation}\label{delta-form}
{\cal D}(\vp)=\d(\vp)d\vp\,,        
\end{equation}
referred  to as {\it Dirac's delta-form} \cite{Itin:2010wt}.   This is a type of a de Rham current, i.e., a singular differential form \cite{Rham}.  
Since the dimension of $\d(\vp)$ is inverse to the dimension of $\vp$, the 1-form ${\cal D}(\vp)$  is  dimensionless independently from  the different possibilities to attach a physical dimension to the function $\vp$. Moreover, the form ${\cal D}(\vp)$ is the same for different functional representations of the hypesurface. Particularly, we can describe the surface by the equation $\vp(x^i)=0$ as well as by an infinite family of  equations $f(\vp(x^i))=0$, with an arbitrary monotonic function $f$. For such different functional representations, we will have different  delta-functions with the same delta-form (up to the leading sign). Indeed, for a monotonic $\psi=f(\vp)$, 
\begin{equation}\label{delta-form-x}
{\cal D}(\psi)=\d(f(\vp))d(f(\vp))=\frac 1{|f'|}\d(\vp)  f'd\vp=\pm {\cal D}(\vp)\,.        
\end{equation}
Similarly to the known representation of the Dirac delta-function as a formal derivative of the step-function, 
the Dirac delta-form can be viewed   as the exterior differential of the step-function $u(\vp)$
 \begin{equation}\label{step}
\D(\vp)=d(u(\vp))\,.
\end{equation}
In a coordinate basis, (\ref{delta-form}) reads 
\begin{equation}\label{delta-coord}
\D(\vp)=\d(\vp)\vp_{,i}dx^i\,.
\end{equation}
We observe that in tensorial equations (written with explicit indices), the term $\d(\vp)\vp_{,i}$ has an invariant meaning, but not $\d(\vp)$ itself.  The delta-function $\d(\vp)$, when applied to the boundary function $\vp(x^i)$, is not well-defined and can come with  different physical dimensions.  

Now we are able  to deal with  a singular 3-form of an electric surface current  in a 4-dimensional
space.  For a given surface $\vp(x)=0$, we define a twisted form of an electric surface  current as 
\begin{equation}\label{delta-coord-1}
^{(sur)}\!\J=L\wedge\D(\vp)=L\d(\vp)\wedge d\vp\,,
\end{equation}
where $L$ is an arbitrary smooth twisted regular 2-form.  This definition  guarantees explicitly the principal properties of the surface current:
\begin{itemize}
\item[(i)] The surface electric current in the 4D-manifold is given as a 3-form.  Hence it is suitable  for being substituted into the inhomogeneous field equation (\ref{curr-1}). 

\item[(ii)]  The current is localized only on the hypersurface $\vp=0$. This property is provided by the factor $\d(\vp)$. Thus, for $\vp\ne 0$ (outside the surface), the current  vanishes.

\item[(iii)]  The current is tangential to the hypersurface. This property is guaranteed by the factor $d(\vp)$. The tangential relation is represented in a metric-free form as 
\begin{equation}\label{surf-identity}
^{(sur)}\!\J\wedge d\vp=0\,.
\end{equation}

\item[(iv)] The absolute dimension  of the 2-form $L$ and of the 3-form $^{(sur)}\!\J$ are the same, namely the dimension of a charge.

\item[(v)] The surface current is independent of the (uncontrolled) choice of the boundary function $\vp(x^i)$.
\end{itemize}

To have a coordinate expression for the surface current (\ref{delta-coord-1}), we recall   the usual representation of a 2-form
\begin{equation}\label{L-coord}
L=\frac 12 \,L_{ij}dx^i\wedge dx^j\,.
\end{equation}
When it is combined   with (\ref{delta-coord}), we have 
\begin{equation}\label{J-coord}
^{(sur)}\!\J=\frac 12 \,L_{ij}\vp_{,k}\d(\vp)dx^i\wedge dx^j\wedge dx^k\,.
\end{equation}
Comparing  with the ordinary representation of the 3-forms (\ref{curr}), we obtain
\begin{equation}\label{J-coord1}
^{(sur)}\!J_{ijk}= L_{[ij}\vp_{,k]}\d(\vp)\,,
\end{equation}
where the antisymmetrization of the indices is denoted by  square parentheses. 
The corresponding dual tensor $J^i=\ep^{ijkl}J_{jkl}$  is expressed by  Levi-Civita dual of the tensor $L_{ij}$
\begin{equation}\label{J-coord2}
^{(sur)}\!J^i= L^{ij}\vp_{,j}\d(\vp)\,,\qquad \mbox{where} \qquad L^{ij}=(1/2)\ep^{ijkl}L_{kl}\,.
\end{equation}

\section{Boundary conditions from the differential Maxwell equations}
Let us now derive the  covariant  boundary conditions directly from the Maxwell equations. 
\subsection{Homogeneous conditions}
The discontinuous  2-form $F$ can be written with the use  of the  Heaviside step-function $u(t)$,
\begin{equation}\label{Hevi}
u(\vp)=\left\{ \begin{array}{ll}
         1 & \mbox{if $\,\,\vp > 0$}\,,\nonumber\\
        0 & \mbox{if $\,\, \vp < 0$}\,.\end{array} \right.
\end{equation}
The value $u(0)$ is not prescribed, but assumed to be final.
 We write the 2-form of the field strength  in the whole space as 
\begin{equation}\label{F-dist}
F(x^i)={}^{(+)}F(x^i) u\big(\vp\big)+ {}^{(-)}F(x^i) u\big(-\vp\big)=\left\{ \begin{array}{ll}
         {}^{(+)}F(x^i) & \mbox{if $\,\, \vp(x^i)> 0$}\,,\\
\\
       {}^{(-)}F(x^i) & \mbox{if $\,\, \vp(x^i) < 0$}\,.\end{array} \right.\,
\end{equation}
Calculating the exterior derivative of both sides of (\ref{F-dist}), we have
\begin{eqnarray}\label{F-dist1}
dF&=&d\Big({}^{(+)}F\Big) u\big(\vp\big)+ d\Big({}^{(-)}F\Big) u\big(-\vp\big)\nonumber\\
&&+{}^{(+)}F \wedge d\Big(u\big(\vp\big)\Big)+{}^{(-)}F  \wedge d\Big(u\big(-\vp\big)\Big)
\,.
\end{eqnarray}
The right-hand side of this equation includes terms of two different types --- the regular (bounded) terms, given in the first line, and the singular (unbounded) terms, listed in the second line. The left-hand-side of (\ref{F-dist1}) is also singular. First, we observe that due to the local equations, 
\begin{equation}\label{F-dist-2}
d\Big({}^{(-)}F\Big)=d\Big({}^{(+)}F\Big)=0\,.
\end{equation}
Thus,  the regular terms are equal to zero. 
A precise meaning of the formal differential equations of the  singular terms can be given as soon as  the integrations are performed on both sides of the equation over some proper regions. Consider a piece $\Sigma$ of a 3D-surface which is transversal to the surface $\vp(x^i)=0$. Let the integral of a singular 3-form $dF$ be {\it defined} due to Stokes' theorem as 
\begin{equation}\label{F-dist1x}
\int_\Sigma dF:=\int_{\partial \Sigma} F\,.  
\end{equation}
Observe that the right-hand-side includes an integration of a non-singular form. Consequently, the left-hand side of this equation is an unambiguous quantity.    
Now we can describe  (\ref{F-dist1}) as an equality between the integrands of the well-defined integrals
\begin{eqnarray}\label{F-dist1xx}
\int_\Sigma dF&=&\int_\Sigma{}^{(+)}F \wedge d\Big(u\big(\vp\big)\Big)+\int_\Sigma{}^{(-)}F  \wedge d\Big(u\big(-\vp\big)\Big)\,.
\end{eqnarray}
Let us  assume  that  the homogeneous  Maxwell equation, $dF=0$, holds in the whole space also for the singular forms. 
Thus, we are left with 
\begin{equation}\label{F-dist-3}
\int_\Sigma {}^{(+)}F \wedge d\Big(u\big(\vp\big)\Big)+\int_\Sigma {}^{(-)}F  \wedge d\Big(u\big(-\vp\big)\Big)=0\,.
\end{equation}
Replacing the exterior derivative of the Heaviside step-function by  Dirac's delta-form, we have
\begin{equation}\label{F-dist-4}
\int_\Sigma {}^{(+)}F \wedge \delta\big(\vp\big)d\vp -\int_\Sigma{}^{(-)}F  \wedge \delta\big(-\vp\big)d\vp=0\,.
\end{equation}
Since delta-function is even, $ \delta\big(-\vp\big)= \delta\big(\vp\big)$, we can rewrite it as 
\begin{equation}\label{F-dist-5}
\int_\Sigma\left[\,{}^{(-)}F(x^i)-{}^{(+)}F(x^i)\right]\wedge    \delta\big(\vp\big)d\vp=0\,.
\end{equation}
Because of the factor $\delta\big(\vp\big)$,  we can replace the arguments of the $F$'s by their limiting values similarly as in  (\ref{bound-04}). Consequently, 
\begin{equation}\label{F-dist-6}
 \int_\Sigma \big[ F\big] \wedge\D \big(\vp\big)=0\,.
\end{equation}
This equation must hold for an arbitrary surface $\Sigma$.  Thus, we have the homogeneous boundary condition in a compact covariant  metric-free form as 
\begin{equation}\label{F-dist-7}
 \big[ F\big] \wedge d\vp=0\,.
\end{equation}
Here, the coordinates in $F=F(x^i)$ are assumed to be constrained by $\vp(x^i)=0$. 
\subsection{Inhomogeneous conditions.}
 Analogously  to the field strength $F$, the excitation field $\H$ in the whole space is written as a combination 
\begin{equation}\label{H-dist}
\H={}^{(+)}\H u\big(\vp\big)+ {}^{(-)}\H u\big(-\vp\big)\,.
\end{equation}
We assume the generic electric current to be  a sum of  the surface and bulk currents  
\begin{equation}\label{J-dist}
J={}^{(sur)}J+{}^{(bulk)}J\,.
\end{equation}
We also assume that the bulk current can be decomposed as
\begin{equation}\label{J-dist-1}
{}^{(bulk)}J={}^{(+)}J u\big(\vp\big)+ {}^{(-)}J u\big(-\vp\big)\,.
\end{equation}
Calculating the exterior derivative of both sides of (\ref{H-dist}), we have  
\begin{eqnarray}
dH&=&d\Big({}^{(+)}H\Big) u\big(\vp\big)+d\Big({}^{(-)}H\Big) u\big(-\vp\big)\nonumber\\&&+{}^{(+)}H \wedge d\Big(u\big(\vp\big)\Big)
+{}^{(-)}H\wedge d\Big(u\big(-\vp\big)\Big)\,.
\end{eqnarray}
Here, the right-hand-side also includes the regular terms, which we list in the first line, and the singular terms listed in the second line. The regular terms are compensated by the bulk current due to the  local (point-wise) equations
\begin{equation}\label{H-reg}
d\Big({}^{(\pm)}H\Big)={}^{(\pm)}J \,.
\end{equation}
Consequently, we are left with the equation 
\begin{equation}\label{equat-dif}
{}^{(+)}H \wedge d\Big(u\big(\vp\big)\Big)+{}^{(-)}H\wedge d\Big(u\big(-\vp\big)\Big)={}^{(sur)}J\,.
\end{equation}
The precise meaning of this equation is expressed by the integral equation between the singular terms
\begin{equation}\label{equat}
\int_\Sigma{}^{(+)}H \wedge d\Big(u\big(\vp\big)\Big)+\int_\Sigma{}^{(-)}H\wedge d\Big(u\big(-\vp\big)\Big)=\int_\Sigma{}^{(sur)}J\,.
\end{equation}
Recall that $\Sigma$ is a small connected part of a 3D-surface which is transversal to the boundary $\phi(x^i)=0$.  

Consequently, the inhomogeneous boundary condition reads
\begin{equation}\label{inh-cond}
  \big[ H\big] \wedge\D \big(\vp\big) ={}^{(sur)}J\,.
\end{equation}
Using (\ref{delta-coord-1}) we can rewrite this equation in terms of the surface 2-form $L$,
\begin{equation}\label{inh-cond0}
\int_\Sigma \Big(\big[ H\big]-L\Big) \wedge\D \big(\vp\big) =0\,.
\end{equation}
This equation must hold for an arbitrary choice of the integration domain $\Sigma$, thus the second boundary condition is:
\begin{equation}\label{inh-cond1x}
 \Big(\big[ H\big]-L\Big) \wedge\D \big(\vp\big) =0\,.
\end{equation}
When the equation $\vp(x^i)=0$ is substituted here, we arrive at the  non-singular condition
\begin{equation}\label{inh-cond1}
 \Big(\big[ H\big]-L\Big) \wedge d\vp =0\,.
\end{equation}

\section{Boundary conditions from the integral Maxwell equations}
In this section, we derive the boundary conditions from the integral Maxwell equations. These  equations allow to consider  non-differentiable and even non-continuous  fields.
 Notice a simple fact: When we integrate  the 3-forms $dF$, $d\H$ and $\J$, the domains of integration must be chosen to be a part of a 3-dimensional surfaces embedded in a  4-dimensional spacetime manifold. Similarly, the integrals of the 2-forms $F$ and $\H$ can be considered only over 2-dimensional  domains. In the literature, see for instance \cite{nasa},  we can find integrals of the corresponding tensorial quantities taken over 4-dimensional regions.  Such a procedure is certainly  not invariant. If we use differential forms instead of tensors, we can determine the appropriate  domains of integration.

\subsection{ Field strength jump condition} 
In the integral form, the homogeneous Maxwell equation $dF=0$ is written as 
\begin{equation}\label{int1}
\int_{C} F=0\,,  
\end{equation}
where $C$ is a 2-dimensional  boundary of some connected  3-dimensional domain $\Sigma$, i.e., $C=\partial \Sigma$.  Consequently, $\partial C =0$.

Let $\Sigma$ denote a  part of a  3-dimensional surface (Fig 1.) that is transversal to the boundary $\vp(x^i)=0$ and bounded by the 3-surfaces $\vp(x^i)=+\epsilon$ and $\vp(x^i)=-\epsilon$.  Geometrically it means that the boundary $\vp(x^i)=0$ divides $\Sigma$ into two connected disjoint pieces $\Sigma_+$ and $\Sigma_-$. 
The 2-form  $F$ is assumed to be smooth in the domains $\Sigma_+$ and $\Sigma_-$. In the whole domain $\Sigma$, it is discontinuous, but finite (bounded). 

Denote the parts of the boundary $\partial \Sigma$ lying on  the surfaces $\vp(x^i)=\pm\epsilon$ by $C_{+\epsilon}$ and $C_{-\epsilon}$, respectively. 
Let $\widetilde{C}$ denotes the remaining part of $\partial\Sigma$, 
see Fig. ~1. 
Due to (\ref{int1}), we have
\begin{equation}\label{int2}
\int_{C_{+\epsilon}}{} ^{(+)} F+\int_{C_{-\epsilon}}{}^{(-)}F+\int_{\widetilde{C}} F=0\,.  
\end{equation}
When the limit $\epsilon\to 0$ is taken, the third  integral goes to zero, since  the domain of integration ${\tilde{C}}$ approaches  zero. The domains  $C_{+\epsilon}$ and $C_{-\epsilon}$ approach the same domain $C_0$ lying in the 3-dimensional surface $\vp=0$. 
Due to the opposite  orientations of the domains $C_{+\epsilon}$ and $C_{-\epsilon}$, we are left   with 
\begin{equation}\label{int3}
\int_{C_0}\big[ F\big]=0\,.  
\end{equation}
Since $C_0$ is an arbitrary domain embedded in the surface with a constant value of the function $\vp$, the recent equation yields 
\begin{equation}\label{int6}
 \big[ F\big] \wedge d\vp=0\,.
\end{equation}

\subsection{ Excitation jump condition} 
The inhomogeneous Maxwell equation $d\H=\J$ can be written in the integral form as
 \begin{equation}\label{int7}
\int_{\partial \Sigma}\H=\int_{\Sigma}\J\,.  
\end{equation}
Here $\Sigma$ is an arbitrary 3-dimensional domain bounded by the closed 2-dimensional surface $\partial \Sigma$. 
Using the same domain as above, we have 
 \begin{equation}\label{int7x}
\int_{\partial \Sigma}\H=\int_{C_{+\epsilon}}\!\!\!\!{}^{(+)} H+\int_{C_{-\epsilon}}\!\!\!\!{}^{(-)}\H+\int_{\widetilde{C}}\!\! \H\,.  
\end{equation}
\parbox [t ]{0.45\textwidth }{
\includegraphics[width=1\linewidth]{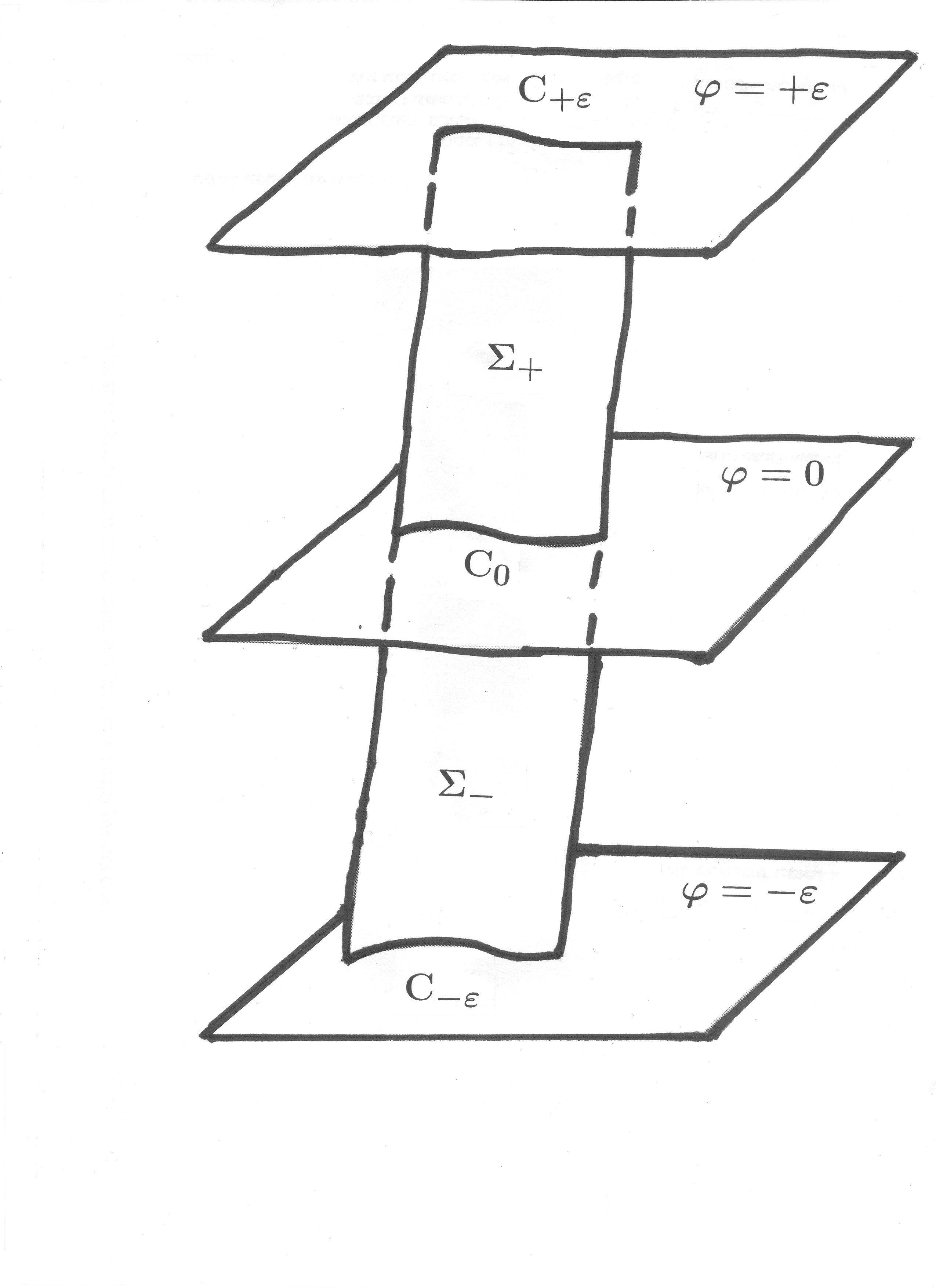}}
\hfill
\parbox [t ]{0.45\textwidth }{
Fig.~1. Here $\Sigma$ is a  part of a  3-dimensional surface, which is transversal to the boundary $\vp(x^i)=0$ and bounded by the 3-surfaces $\vp(x^i)=+\epsilon$ and $\vp(x^i)=-\epsilon$.  The boundary $\vp(x^i)=0$ divides $\Sigma$ into two connected disjoint pieces $\Sigma_+$ and $\Sigma_-$. 
The parts of the boundary $\partial \Sigma$ lying on  the surfaces $\vp(x^i)=\pm\epsilon$ are denoted by $C_{+\epsilon}$ and $C_{-\epsilon}$ correspondingly.  $\widetilde{C}$ denotes the remaining part of $\partial\Sigma$. }


When the limit $\epsilon\to 0$ is considered, the third integral trends to zero, while  the domains $C_+$ and  $C_-$ go to the same limit $C_0$. Hence, we are left with
 \begin{equation}\label{int8}
\int_{\partial \Sigma}\H=\int_{C_0}\big[\H\big] \,.  
\end{equation}
The total current containing in the domain $\Sigma$ consists from the bulk and surface current
 \begin{equation}\label{int9}
\int_{\Sigma}\J=\int_{\Sigma}\bigg({}^{({\rm bulk})}\J+ {}^{({\rm surf})}\J\bigg)\,.  
\end{equation}
In the limit $\epsilon\to 0$,  the bulk current vanishes, thus the total charge is 
 \begin{equation}\label{int10}
\int_{\Sigma}\J= \int_{\Sigma}L\wedge\D\vp=\int_{C_0}L\,.  
\end{equation}
Equating (\ref{int8}) and  (\ref{int10}), we obtain
\begin{equation}\label{int11}
\int_{C_0}\bigg(\big[H\big]-L\bigg)=0\,.  
\end{equation}
This integral equation is of the same type as (\ref{int3}). Due to the arbitrariness of the domains $C_0$, it yields the boundary condition
\begin{equation}\label{int12}
\bigg(\big[H\big]-L\bigg)\wedge d\vp=0\,.  
\end{equation}
\section{Analysis of boundary conditions}
In this section, we analyze the boundary conditions derived above:
\begin{equation}\label{bound-form}
 \boxed{\big[F\big] \wedge d \vp =0\,,\qquad  \Big(\big[ H\big]-L\Big) \wedge d\vp =0\,.}
\end{equation}
\subsection{Boundary conditions in tensorial form}
Although the differential form representation is very  compact,  the tensorial representation is also useful, particularly for a comparison with the literature. 
 Let a part of a boundary be embedded in a domain with the coordinates $x^i,\, i=0,1,2,3$. 
In these coordinates, the differential of the boundary function is expressed as 
$d \vp=\vp_{,k}dx^k$. The coordinate representations of the fields  are given in (\ref{forms}).
The first equation of (\ref{bound-form}) can be rewritten  as
 \begin{equation}\label{bound-form1x}
 \big[F_{ij}\big] \vp_{,k}dx^i\wedge dx^j \wedge dx^k =0\,.
\end{equation}
We can even remove the basis 1-forms, but we must preserve the anti-symmetriza\-tion in the indices. This can be done by using  the permutation pseudo-tensor. 
So we write the first boundary condition as
 \begin{equation}\label{bound-form1}
\ep^{ijkl} \big[F_{ij}\big] \vp_{,k}=0\,,
\end{equation}

The second equation of (\ref{bound-form}) is of the same structure as the first one. We only need a coordinate representation of the 2-form $L$. It is given  in (\ref{L-coord}).
Hence, the second equation becomes
\begin{equation}\label{bound-form3}
 \Big(\big[H^{ij}\big]-L^{ij}\Big) \vp_{,m}\ep_{ijkl}dx^k\wedge dx^l \wedge dx^m =0\,.
\end{equation}
The basis forms can also  be removed here, but the anti-symmetrization must be preserved, i.e., an additional permutation pseudo-tensor must be used.  
Thus, we arrive at the tensorial representation of  the boundary conditions, 
\begin{equation}\label{bound-form4}
\boxed{\ep^{ijkl}\big[F_{jk}\big] \vp_{,l}=0\,,\qquad \Big(\big[H^{ij}\big]-L^{ij}\Big) \vp_{,j}=0\,.}
\end{equation}

\subsection{Spacetime decomposition}
Let us fix the coordinate system and decompose the 4-dimensional tensorial fields into 3-dimensional vector fields. The decomposition of the fields is   given in  (\ref{vec-ten3}--\ref{vec-ten2}). 
Denote the components of the 4D gradient $\vp_{,i}$ by 
\begin{equation}\label{comp1}
\vp_{,0}=n_0\,,\qquad \vp_{,\a}=n_\a(=\mathbf n)\,.
\end{equation}
Recall that the Greek indices  run over 1,2, and 3.  


\vspace{0.5 cm}

\noindent {\bf Homogeneous conditions:}
For $i=0$, we rewrite (\ref{bound-form4}) as 
\begin{equation}\label{decomp1}
\ep^{\a\b\g}\big[F_{\a\b}\big] \vp_{,\g}=0\,,
\end{equation}
where the standard identification $\ep^{0\a\b\g}=\ep^{\a\b\g}$ is used.
Using (\ref{vec-ten1}), we have 
\begin{equation}\label{decomp2}
\big[B^{\a}\big] \vp_{,\a}=0\,.
\end{equation}
In  ordinary vectorial notation, it can be rewritten as 
\begin{equation}\label{decomp3}
\big[{\mathbf B}\big] \cdot{\mathbf n}=0\,.
\end{equation}
Recall that the dot-product used here is independent of the metric.  It means a contraction of a vector with a covector, rather than a scalar product of two vectors.   
For a spatial value of the index $i$, say $i=\a$, we rewrite (\ref{bound-form4}) as
\begin{equation}\label{decomp4}
\ep^{\a0\b\g}\big[F_{0\b}\big]\vp_{,\g}+\ep^{\a\b0\g}\big[F_{\b0}\big]\vp_{,\g}+\ep^{\a\b\g0}\big[F_{\b\g}\big]\vp_{,0}=0\,.
\end{equation}
Using (\ref{vec-ten3}), we have 
\begin{equation}\label{decomp5}
\ep^{\a\b\g}\big[E_{\b}\big]n_{\g}-\big[B^\a\big]n_{0}=0\,.
\end{equation}
In  vectorial notation, it is rewritten as 
\begin{equation}\label{decomp6}
\big[{\mathbf E}\big]\times{\mathbf n}-\big[{\mathbf B}\big]n_{0}=0\,.
\end{equation}

\vspace{0.5 cm}

\noindent {\bf Inhomogeneous conditions:}
The structure of the second equation of (\ref{bound-form4}) is similar to the first one. Thus,  for $i=0$, we have
\begin{equation}\label{decomp7}
\big[H^{0\a}\big]\vp_{,\a}=L^{0\a} \vp_{,\a}\,,
\end{equation}
or, equivalently,
 \begin{equation}\label{decomp8}
\big[D^{\a}\big]n_{\a}=-L^{0\a} n_{\a}\,.
\end{equation}
For the spatial value of the index, say  $i=\a$, we have 
\begin{equation}\label{decomp9}
\Big(\big[H^{\a\b}\big]-L^{\a\b} \Big)\vp_{,\b}+\Big(\big[H^{\a0}\big]-L^{\a0} \Big)\vp_{,0}=0\,.
\end{equation}
In terms of the 3D vectors, it is rewritten as
\begin{equation}\label{decomp10}
\ep^{\a\b\g}\big[H_\g\big]n_{\b}+\big[D^{\a}\big]n_{0}=L^{\a\b} n_{\b}+L^{\a0} n_{0}\,.
\end{equation}
Comparing with the inhomogeneous  static boundary  conditions  (\ref{bound-1a}), we rewrite (\ref{decomp8}) and (\ref{decomp10}) in  vectorial form as
\begin{equation}\label{decomp11}
\big[{\mathbf D}\big] \cdot{\mathbf n}=\sigma\,, \qquad \big[{\mathbf H}\big]\times{\mathbf n}+\big[{\mathbf D}\big]n_{0}={\mathbf K}\,.
\end{equation}
Here the surface charge scalar and the surface current vector are defined as 
\begin{equation}\label{decomp12}
\sigma=-L^{0\a} n_{\a}\,, \qquad K^\a=-L^{\a\b} n_{\b}-L^{\a0}n_{0} \,.
\end{equation}

\subsection{Independent conditions} 
The number of independent boundary conditions in electromagnetism is discussed in literature, see \cite{Yeh},\cite{Gur} and the references given therein.  
In the differential form notations, the boundary conditions are written as
\begin{equation}\label{ind1}
 C_1=\big[F\big] \wedge \D \big(\vp\big) =0\,,\qquad  C_2=\Big(\big[ H\big]-L\Big) \wedge \D \big(\vp\big) =0\,.
\end{equation}
We have here two 3-forms $C_1$ and $C_2$ which are equal to zero, i.e., there are eight conditions linear in the field jumps. These conditions, however, are not independent. Indeed, one can easily observe two linear constraints
\begin{equation}\label{const}
C_1\wedge d\vp=C_2\wedge d\vp=0\,.
\end{equation}
 Thus, in the covariant differential form description, only six  linear independent conditions remain.

In the vectorial representation, we have a system of eight  equations 
\begin{eqnarray}\label{vect-cond}
\big[{\mathbf B}\big] \cdot{\mathbf n}=0\,,&& \qquad \big[{\mathbf D}\big] \cdot{\mathbf n}=\sigma \,,\nonumber\\
\big[{\mathbf E}\big]\times{\mathbf n}-\big[{\mathbf B}\big]n_{0}=0\,,&& \qquad  \big[{\mathbf H}\big]\times{\mathbf n}+\big[{\mathbf D}\big]n_{0}={\mathbf K}\,.
\end{eqnarray}
Observe here two different possibilities:

\vspace{0.2 cm}

\noindent {\bf Non-static boundary:} For a moving boundary with $n_0=\partial \vp/\partial x^0\ne 0$, two scalar equations in the first line of (\ref{vect-cond}) are the results of the vector equations given in the second line. 
To see these relations, it is enough to apply the dot-multiplication of the vector equations with the vector ${\mathbf n}$  and use the charge conservation law. 
Thus, we are left with only six independent conditions: 
\begin{eqnarray}\label{mov-vect-cond}
\big[{\mathbf E}\big]\times{\mathbf n}-\big[{\mathbf B}\big]n_{0}=0\,,&& \qquad  \big[{\mathbf H}\big]\times{\mathbf n}
+\big[{\mathbf D}\big]n_{0}={\mathbf K}\,.
\end{eqnarray}
This is in an agreement with the covariant consideration given above. 

\vspace{0.2 cm}

\noindent{\bf Static boundary.} In this case when $\vp$ is independent of the $x^0$ coordinate, the scalar $n_0=0$, thus we are left with eight boundary conditions 
   \begin{eqnarray}\label{st-vect-cond1}
\big[{\mathbf B}\big] \cdot{\mathbf n}=0\,,&& \qquad \big[{\mathbf D}\big] \cdot{\mathbf n}=\sigma \,,
\label{st-vect-cond2} \\
\big[{\mathbf E}\big]\times{\mathbf n}=0\,,&& \qquad  \big[{\mathbf H}\big]\times{\mathbf n}={\mathbf K}\,.
\end{eqnarray}
Note that we did not use any constitutive relation so far. In this situation, four vector fields are completely independent, hence also the equations 
(\ref{st-vect-cond1}-\ref{st-vect-cond2}) seem to be  independent. In fact, this is not the case. Indeed, in (\ref{st-vect-cond1}) we have two independent scalar conditions. As for the vector equations given in (\ref{st-vect-cond2}), not all of them are independent. Denoting  ${\bf A}_1=\big[{\mathbf E}\big]\times{\mathbf n}$, we can easily observe a linear relation ${\bf A}_1\cdot{\mathbf n}=0$. Similarly, for the second vector ${\bf A}_2=\big[{\mathbf H}\big]\times{\mathbf n}-{\mathbf K}$, we have an identity ${\bf A}_2\cdot{\mathbf n}=0$, providing the surface current being tangential to the surface ${\mathbf K}\cdot{\mathbf n}=0$. 
Thus, we have the same six boundary conditions  as in the non-static case. 
\section{Energy transfer}
\subsection{Metric-free energy-momentum current}
In  relativistic physics, the energy of a field together with its momentum and its stress tensor form a four-dimensional energy-momentum tensor $T^j{}_i$. Due to the GR paradigm, the conservation low for this tensor must be expressed by use of the covariant derivative $T^j{}_{i;j}=0$. Although this equation is ordinarily used as the GR representation of the conservation law,  it does not provide an invariant integral conservation law. In the differential form representation, the energy-momentum current $\Sigma_i$ is considered instead of the tensor $T^j{}_i$. The relation between the tensor $T^j{}_i$ and the form $\Sigma_i$ is provided by use of the volume element and does not require the metric structure \cite{Itin:2003jp}. $\Sigma_i$ is a covector valued 3-form  that obeys the following conservation law:
\begin{equation}\label{en1}
 d\Sigma_i =\F_i\,,
\end{equation}
where $\F_i$ is a covector valued 4-form of the force density. It should be noted that the equation (\ref{en1}) is not invariant under general coordinate transformations, and the integral formulation is not applicable also in this formulation. 

The same problem appears in the electromagnetism of media. Different definitions of the electromagnetic energy-momentum tensor are known to be applicable in different circumstances, but it is not clear yet from both a theoretical \cite{Ramos:2011mr} and an experimental \cite{Brevik:1979zf} point of view, which of them must be taken as the fundamental one.  For the purpose of the analysis of the boundary conditions, we will apply the definition which can be traced back to Minkowski. 
A corresponding tensor (form) is derived from the electromagnetic Lagrangian by the standard Noether-Hilbert procedure. This fact distinguishes  Minkowski's tensor from the other energy-momentum expressions.

 In the premetric approach, the energy-momentum current of the electromagnetic field is expressed as 
\begin{equation}\label{en2}
 \Sigma_i =\frac 12 \left[\left(e_i\rfloor F\right)\wedge H-\left(e_i\rfloor H\right)\wedge F\right]\,,
\end{equation}
 where the interior multiplication operator $\rfloor$ is applied. 

When the exterior derivative of this expression is evaluated, the corresponding force density $\F_i=d\Sigma_i$ includes the Lorentz force density $\left(e_i\rfloor F\right)\wedge J$ plus an additional force density. This additional force can be expressed via the Lie derivatives of the fields $F$ and $H$ taken relative to the coordinate vectors \cite{birkbook}. In the particularly interesting case of a linear constitutive relation $H^{ij}=(1/2)\chi^{ijkl}F_{kl}$, the additional force is expressed by the first-order derivatives of the constitutive tensor $\chi^{ijkl}$. It should be noted that this additional force is not only a mathematical artifact, but is observed in experiments \cite{Brevik:1979zf}. 

\subsection{Energy-momentum boundary conditions}

To describe the distribution of the energy-momentum current in the case of two media separated by a hypersurface $\vp(x^i)$, we substitute  the definitions (\ref{F-dist}, \ref{H-dist}) of the fields into (\ref{en2}). Using the identities 
\begin{equation}\label{en3}
u(\vp)u(-\vp)=0\,,\qquad [u(\vp)]^2=u(\vp)\,,
\end{equation}
we derive the decomposition of the energy-momentum current as the sum of two independent currents related to the two different media
\begin{equation}\label{en4}
 \Sigma_i = {}^{(-)}\Sigma_iu(\vp)+ {}^{(+)}\Sigma_iu(-\vp)\,.
\end{equation}
Here 
\begin{equation}\label{en5}
 {}^{(-)}\Sigma_i =\frac 12 \left[\left(e_i\rfloor  {}^{(-)}\!F\right)\wedge  {}^{(-)}\!H-\left(e_i\rfloor  {}^{(-)}\!H\right)\wedge  {}^{(-)}\!F\right]\,,
\end{equation}
and similarly for ${}^{(+)}\Sigma_i$. 
Consequently,
\begin{eqnarray}\label{en6}
 d\Sigma_i& =&d\left({}^{(-)}\Sigma_i\right) u(\vp)+ d\left({}^{(+)}\Sigma_i\right)u(-\vp)+\left({}^{(-)}\Sigma_i-{}^{(+)}\Sigma_i\right)\wedge \D(\vp)\nonumber\\
& =&{}^{(-)}\!\F_i u(\vp)+{}^{(+)}\!\F_i u(-\vp)+\left({}^{(-)}\Sigma_i-{}^{(+)}\Sigma_i\right)\wedge \D(\vp)
\,.
\end{eqnarray}
Thus, the jump of the energy-momentum current emerges as  an additional force acting on the boundary hypersurface
\begin{equation}\label{en7}
{}^{(add)}\F_i=\Big[\Sigma_i\Big]\wedge \D(\vp)=\Big[\Sigma_i\Big]\wedge \d(\vp)d\vp\,.
\end{equation}
This force is localized on the boundary hypersurface and is proportional to the
4D-gradient $d\vp$. 
\subsection{Tensorial description of the energy-momentum jump}
Recall that the considerations above do not require any geometrical structure on the manifold. In order to deal  with the ordinary notion of the energy-momentum tensor, we need a volume element to be defined on the manifold. Conventionally,  this quantity is defined by use of the metric tensor $vol\sim \sqrt{-g}$. Instead, we will consider the volume element $vol$ as   a fundamental structure without any relation to the metric. Notice that $vol$ must be a smooth non-vanishing  twisted 4-form with the support on the whole manifold. 

When a volume element is defined, the covector-valued 4-form of the force density is expressed   as 
\begin{equation}\label{ten1}
\F_i=f_i \,vol\,.
\end{equation}
Here $f_i$ is the ordinary (non-twisted) covector. 

The twisted 3-form of  energy-momentum current is expressed  by use of the contraction (inner product) operator $e_i\rfloor$.  
\begin{equation}\label{ten2}
\Sigma_i=-T_i{}^j \,e_j\rfloor vol\,.
\end{equation}
For a coordinate frame $e_i=\partial/\partial x^i$, it is the standard partial derivative operator. Recall the known identity $e_i\rfloor dx^j=\d^j_i$. Substituting (\ref{ten1}, \ref{ten2}) into (\ref{en7}), we obtain
\begin{equation}\label{ten3}
{}^{(add)}f_i\, vol=-\Big[T_i{}^j\Big]\vp_m \d(\vp) \Big(e_j\rfloor vol\wedge dx^m\Big)=\Big[T_i{}^j\Big]\vp_j \d(\vp) vol\,.
\end{equation}
Here, we used an identity $\left(e_j\rfloor vol\right)\wedge dx^m=\d^m_jvol$. From (\ref{ten3}), we have the energy-momentum jump conditions in the tensorial representation  
\begin{equation}\label{ten4}
{}^{(add)}f_i=\Big[T_i{}^j\Big]\vp_j \d(\vp)\,.
\end{equation}
Up to a normalization factor, this result coincides with the one given in \cite{nasa}.  
The space-time decomposed form of this equation can be also found in \cite{nasa}.  

The results of this section are independent of the metric and of the volume element structures, it is clear from  the differential form representation (\ref{en7}). Although the volume element was used for the definitions of the vector $f_i$ and of the tensor $T_i{}^j$, the relation (\ref{ten4}) is independent of the specific choice of the volume form. Moreover, every two volume elements  differ only  by a scalar factor which  cancels on both sides of   (\ref{ten4}). 

\section{Example: Moving boundary in an isotropic medium}
As a simple application of the covariant jump conditions derived above, we consider the following question:

{\it Let on both sides of the hypersurface $\vp=0$ be the same medium. Is it possible to have a jump of  the field even in the absence of  surface charges and currents?}

We will consider a  moving  surface, i.e. one described    in a chosen system of coordinates by the equation 
 $\vp(x^i)=0$ with the derivatives 
\begin{equation}\label{ex0}
\vp_{,0}=n_0\,,\qquad  \vp_{,\a}=n_\a\,.
\end{equation}
As it was described above, the function $\vp$ is defined up to the action  of an arbitrary monotonic function $f(\vp)$. In particular, we can multiply $\vp$ by an arbitrary constant. Thus, we can assume the spatial part $n_i$ to be dimensionless. Consequently,   $n_0$ has the dimension of  velocity. In contrast to the motion of a point described by a velocity vector, we need only a scalar in order to describe the motion of a hypersurface. 
 As an example, a flat uniformly moving  hyperplane will be described by a linear equation
\begin{equation}\label{ex1}
\vp=n_\a x^\a-n_0t=0\,
\end{equation}
with a constant  4D-covector $n_i=(n_0,n_\a)$ and a constant scalar speed  $n_0$.  

The notion of an isotropic medium notion can only  be defined in the 3-dimensional sense. It is described by two parameters: the permittivity  $\varepsilon$ and the permeability $\mu$ that appear in the constitutive relations
\begin{equation}\label{ex2}
D^\a=\varepsilon\d^{\a\b}E_\b\,, \qquad \H_\a=\frac 1\mu \d_{\a\b}B^\b  \,.
\end{equation}
We do not need to restrict ourselves to  constant parameters. Hence $\varepsilon$ and $\mu$ can depend on the  position and the time coordinates.  
 Since the definition of the isotropic medium  involves the 3-dimensional metric $\d_{\a\b}=diag(1,1,1)$, we can use it for lowering and raising  indices. So we return to the standard vectorial notation   and write the constitutive relation in the ordinary textbook form 
\begin{equation}\label{ex3}
{\mathbf D}=\varepsilon{\mathbf E}\,,\qquad {\mathbf B}=\mu{\mathbf H}\,.
\end{equation}

Let us first check what we can get from the static boundary conditions. Substituting (\ref{ex3}) into (\ref{bound-1}, \ref{bound-1a}), we have (recall the vector notation ${\mathbf n}=n_\a$) 
\begin{eqnarray}\label{ex4}
\mu{\mathbf n}\cdot\big[\mathbf H\big]=0\,,\qquad&& \big[{\mathbf E}\big]\times{\mathbf n}=0\,,\\
  \label{ex5}
\varepsilon{\mathbf n}\cdot\big[\mathbf E\big]=0\,,\qquad&& \big[{\mathbf H}\big]\times{\mathbf n}=0\,.
  \end{eqnarray}
Using the vector triple product, we get
\begin{equation}\label{ex5}
{\mathbf n}\times\left(\big[{\mathbf E}\big]\times{\mathbf n}\right)=\big[\mathbf E\big] n^2-{\mathbf n}\left({\mathbf n}\cdot\big[\mathbf E\big]\right)     =\big[\mathbf E\big] {\mathbf n}^2\,.
\end{equation}
Due to  (\ref{ex4}), we have $\big[\mathbf E\big] n^2=0$. Since the Euclidean square ${\mathbf n}^2\ne 0$, we obtain that, in the isotropic medium, the jump of the electromagnetic field on a static surface is forbidden. 

Let us now substitute  the constitutive relations (\ref{ex3}) into the dynamical jump  conditions (\ref{vect-cond}). Due to the parametrization (\ref{ex0}), we obtain
\begin{eqnarray}\label{ex6}
\mu\big[{\mathbf H}\big] \cdot{\mathbf n}=0\,,&& \qquad \varepsilon\big[{\mathbf E}\big] \cdot{\mathbf n}=0 \,,\nonumber\\
\big[{\mathbf E}\big]\times{\mathbf n}-\mu\big[{\mathbf H}\big] n_{0}=0\,,&& \qquad  \big[{\mathbf H}\big]\times{\mathbf n}
+\varepsilon\big[{\mathbf E}\big]  n_{0}=0\,.
\end{eqnarray}
Now we have 
\begin{equation}\label{ex7}
{\mathbf n}\times\left(\big[{\mathbf E}\big]\times{\mathbf n}\right) =\big[\mathbf E\big] {\mathbf n}^2\,,
\end{equation}
which is not equal to zero. 
Due to (\ref{ex6}), this expression is evaluated as 
\begin{equation}\label{ex7}
{\mathbf n}\times\left(\big[{\mathbf E}\big]\times{\mathbf n}\right)=\mu  n_0{\mathbf n}\times\big[{\mathbf H}\big]=\mu\varepsilon  n^2_0\big[\mathbf E\big] \,.
\end{equation}
Thus, we obtain a jump condition
\begin{equation}\label{ex8}
\left({\mathbf n}^2-\mu\varepsilon  n^2_0\right)\big[\mathbf E\big] =0\,.
\end{equation}
Consequently, a non-zero jump of the field $\mathbf E$ is admissible on a surface $\vp(x^i)=0$ whose  gradient $\vp_i=(n_0,{\mathbf n})$ satisfies the relation 
\begin{equation}\label{ex9}
{\mathbf n}^2-\mu\varepsilon  n^2_0=0\,.
\end{equation}
The expression on the left-hand-side  here can be viewed as a Lorentzian type norm of the covector $\vp_i$. The constant of the speed of light  in vacuum  in this expression is substituted by the speed of light in the medium $v=n_0/|{\mathbf n}|$
\begin{equation}\label{ex10}
v=\frac 1{\sqrt{\mu\varepsilon}}\,.
\end{equation}
Thus the 4-dimensional metric emerges in  a priori metric-free background due to the special isotropic constitutive relations. 
This metric is Lorentzian if the product $\mu\varepsilon$ positive. This condition holds for the ordinary materials  with $\mu>0, \varepsilon>0$ as well for the recently generated metamaterials with the parameters $\mu<0, \varepsilon<0$. 
When the parameters $\mu, \varepsilon$ are of  different signs, the metric is Euclidean. 
The non-zero jump of the fields is usually described as the shock waves. Note that we derived such type of  behavior from the jump conditions alone, without using the field equations directly and without assuming any wave-type ansatz.

\section{Results and discussion} We derived the covariant boundary conditions in electrodynamics. The resulting formulas as well as the derivation procedures  do not require the metric structure. Thus, these results are valid  on an arbitrary differential manifold. A key point of our consideration is a proper metric-free redefinition of the electromagnetic fields, the electric current and the energy-momentum current. These quantities are represented by singular differential forms and require Dirac's delta-form for their definitions. The metric-free covariant boundary conditions must have a wide area of applications from GR to the electromagnetism of media. Recently, Kurz and Heumann \cite{kurz} applied the premetric covariant boundary conditions numerically to the classical Wilson experiment.

Our derivation was based on the boundary between two different media. However, due to their metric-free form, the results must hold for an arbitrary hypersurface. In particular, in the source-free form, they are applicable to the wave front surface. A simple isotropic example of such a consideration is presented above.
\section*{Acknowledgment}
  I would like to thank Friedrich Hehl, Volker Perlick and Stefan Kurz  for their most fruitful comments. The main part of this paper was represented at ACE2010 workshop at ETH Z\"urich. I acknowledge  the GIF grant No. 1078-107.14/2009 for financial support. 

  
\end{document}